\documentclass[superscriptaddress,aps,prl,showkeys,showpacs,twocolumn,10pt,nofootinbib]{revtex4-2}
\usepackage[OT1,T1]{fontenc}
\usepackage{graphicx}
\usepackage{rotating}
\usepackage{amsmath}
\begin{document}

\title{On Sequential Single-Pion Production in Double-Pionic Fusion}
\date{\today}

\newcommand*{\PITue}{Physikalisches Institut, Eberhard--Karls--Universit\"at 
 T\"ubingen, Auf der Morgenstelle~14, 72076 T\"ubingen, Germany}
\newcommand*{\Kepler}{Kepler Center for Astro and Particle Physics, University
  of T\"ubingen, Auf der Morgenstelle~14, 72076 T\"ubingen, Germany}
\newcommand*{\York}{Department of Physics, University of York, Heslington,
  York,Y010 5DD,UK} 

\author{M. Bashkanov} \affiliation{\York}
\author{H.~Clement}     \affiliation{\PITue}\affiliation{\Kepler}

\newcommand*{\Moscow}{Now at Moscow}

\begin{abstract}
Recently a two-step process has been proposed for the double-pionic fusion to
deuterium $pn (I=0) \to d\pi^+\pi^-$. Its calculation is solely based on total cross
section data for the two sequential single-pion production steps $pn (I=0)\to
pp\pi^-$ followed by $pp \to d\pi^+$. Though this sequential process was aimed
to explain the dibaryon resonance $d^*(2380)$ peak in double-pionic fusion, we
demonstrate that this is not the case. It rather fits to a
possible broad bump at 2.31 GeV in the energy dependence of the $pn \to d\pi^0\pi^0$ reaction, which was recently interpreted as a consequence
of dibaryonic excitations in isoscalar single-pion production.
\end{abstract}


\maketitle

Two-step processes are well-known in nuclear physics and have been studied
there intensively for decades in a variety of nuclear reactions. In general
their cross section is smaller than a competing direct process by an order 
of magnitude. Hence two-step processes are usually important, if the
direct process is suppressed for some reason.

\begin{figure*} 
\centering
\includegraphics[width=1.99\columnwidth]{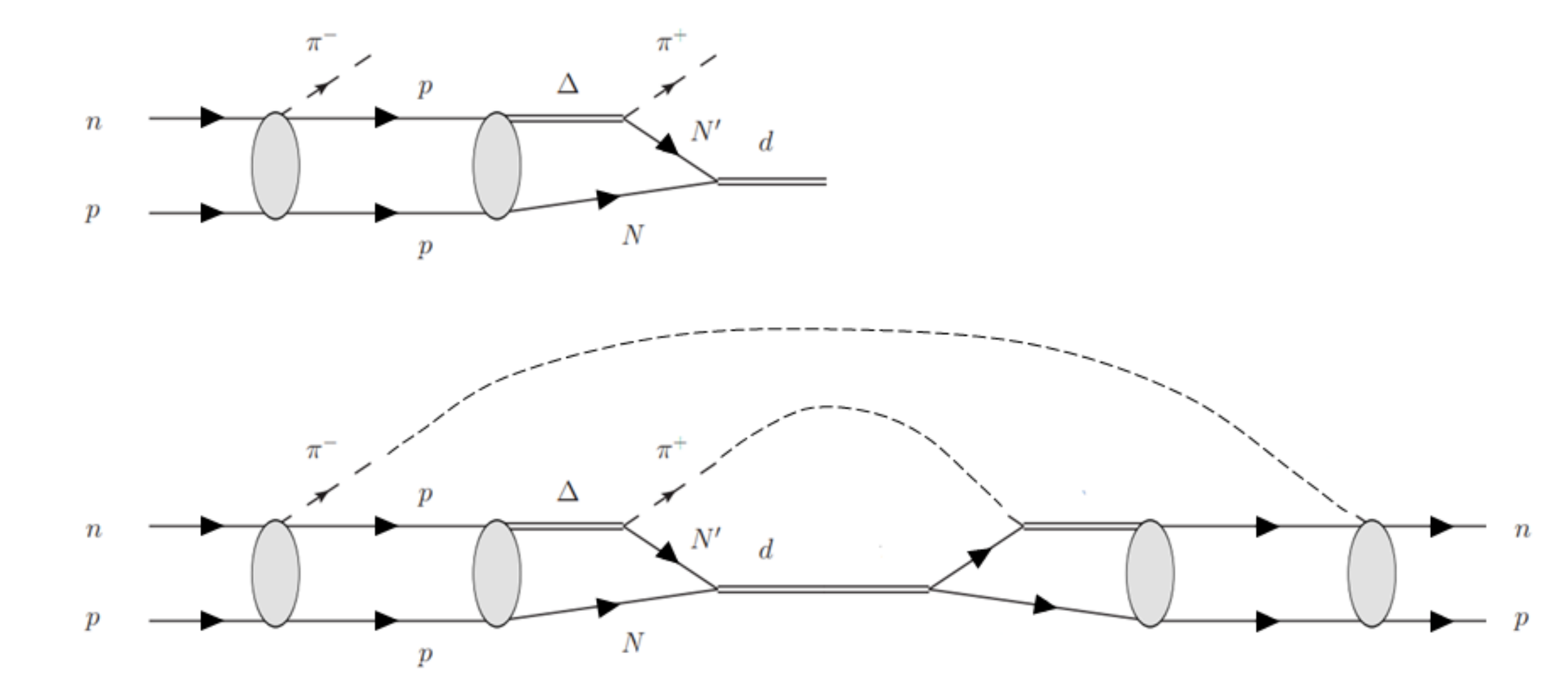}
\caption{\small  
Diagrammatical representation of the two-step process for the $np \to
d\pi^+\pi^-$ reaction (top) suggested in
Refs. \cite{EO,Osetdpi} and its continuation as a four-step process to affect
elastic $np$ scattering as suggested in Ref. \cite{EO} (bottom).
}
\label{fig1}
\end{figure*}

Recently it has been proposed by Molina, Ikeno and Oset \cite{EO,Osetdpi} that a
two-step process in form of two successive single-pion production processes
may happen for the basic double-pionic reaction $pn \to d\pi\pi$. In
particular the isoscalar reaction sequence $pn(I=0) \to (pp)\pi^- \to
(d\pi^+)\pi^-$ has been considered. The two-step process with
explicit $\Delta$ excitation in the second step is
depicted diagrammatically in Fig.~\ref{fig1}(top). In Ref. \cite{EO}
it was argued that this two-step process could produce even a circle in the
Argand plot of a specific partial wave in elastic $np$ scattering, which is a
necessary condition for a true resonance.
It is well known that the chain $pp \to \Delta^+ p \to d\pi^+ \to \Delta^+ p \to
pp$ can reproduce at least part of the loop in the Argand plot
\cite{Anisovich,PK}.
As illustrated in Fig.~\ref{fig1}(bottom) the situation in our case is more
complicated, since the chain $pp \to \Delta^+ p \to d\pi^+ \to \Delta^+ p \to
pp$ is preceeded by the further step $pn(I=0) \to (pp)\pi^-$.

The appealing beauty of the presented formalism is that the total 
cross section in the final $d\pi^+\pi^-$ channel is claimed to be calculable by just the
knowlege of the total cross sections of the participating reactions $pn(I=0)
\to pp\pi^-$ (step-1 reaction) and $pp \to d\pi^+$ (step-2 reaction), which
both have been studied experimentally, phenomenologically and theoretically. The 
drawback of the formalism presented in Ref. \cite{EO}, of course, is that no differential cross sections
can be calculated and that no spin-parity quantum numbers are selected. Hence
one has to be very careful in the interpretation of the results. 

In Ref. \cite{EO} the intention was to present an alternative explanation of the
$d^*(2380)$ dibaryon resonance structure with $I(J^P) = 0(3^+)$ in the $np \to
d \pi^+\pi^-$ reaction by use of the formalism for the sequential single-pion
production. The pole of the $d^*(2380)$ resonance has been identified
at 2.38 GeV both in polarized \cite{np,npfull} and unpolarized \cite{npel}
elastic neutron-proton scattering by use of the full SAID database, albeit the
critical contribution came from the polarization data \cite{np,npfull}. The
$d^*(2380)$ resonance has been observed in  the isoscalar part
of all the various $NN\pi\pi$ channels \cite{MB,d00,d00pol,d+-,pp0-,pn00,pn+-}
exhibiting there a pronounced narrow Lorentzian of width 70 MeV. Hence the
two-step process should undergo such a narrow structure around 2.38 GeV at least in one
of the two participating reactions (In principle one can get a peak also, if one process is rising, whereas the other one is falling). The total cross section of the step-2
reaction, the $pp \to d\pi^+$ reaction, exhibits only a broad resonance 
structure due to the $\Delta(1232)$ excitation \footnote{For recent 
  interpretations see, {\it e.g.}, Refs.~\cite{PK,Niskanen}}. Therefore the desired
structure must be found in the step-1 reaction, the $pn(I=0) \to pp\pi^-$
reaction. The experimental isoscalar cross section exhibits indeed a bump
structure around 2.31 GeV, but again only a broad one, which was fitted in
Ref. \cite{NNpi} by a Gaussian of width 150 MeV. However, by 
increasing the error bars of the WASA data by an order of magnitude and ignoring recent high-precision data from Gatchina~\cite{AS},
Ref. \cite{EO} succeeded to achieve a seemingly alternative  description with
$\chi^2 \ll$ 1 providing now Breit-Wigner shapes peaking at 2.33 - 2.34 GeV
and having a width of 70 - 80 MeV. In Ref. \cite{EO} this could be
achieved only by enlarging the uncertainties of the WASA-at-COSY results
enormously by adding in quadrature a large systematic error arguing
that is due to the neglect of isospin violation in the derivation of the
$pn(I=0) \to pp\pi^-$ data in Ref. \cite{NNpi}. Such a procedure of handling
systematic errors as presented
in Ref. \cite{EO} is by no means justified, since the
isospin violation is not fluctuating randomly from energy point to energy point
and hence does not behave like statistical uncertainties. Therefore it
cannot be added to them. Isospin violation rather affects just the
absolute scale of the isoscalar cross section shifting the data solely in common
up or down in scale. It was also shown in Ref.~\cite{d+-} that isospin violation strongly affects the shape of differential $M_{\pi\pi}$ distributions due to different thresholds for $M_{\pi^0\pi^0}$ and $M_{\pi^+\pi^-}$, see, {\it e.g.}, Ref. \cite{PK}. Hence, inability to correctly reproduce the differential observables would unavoidably lead to an incorrect isospin violation prediction.

\begin{figure} 
\centering
\includegraphics[width=0.99\columnwidth]{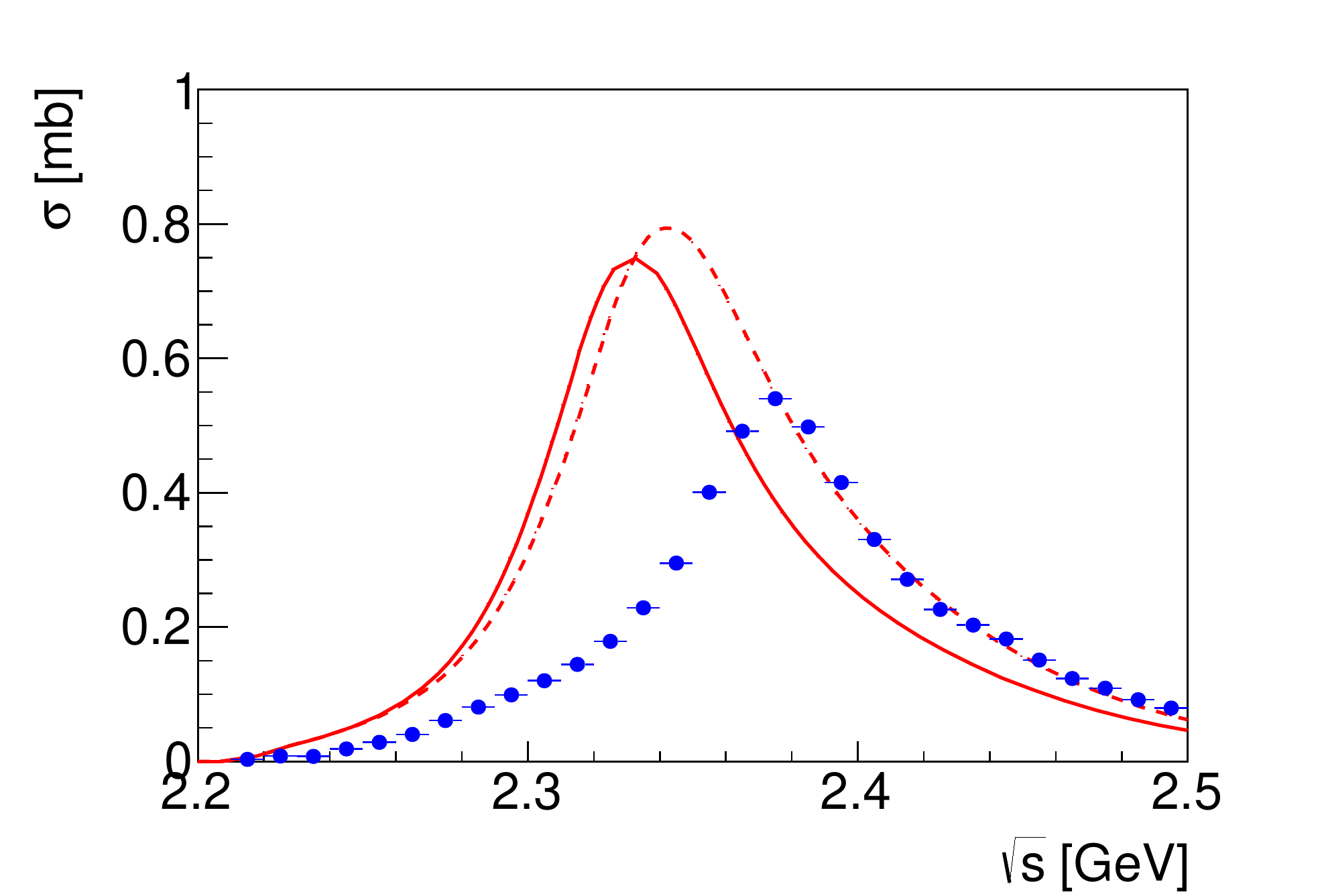}
\caption{\small 
The isoscalar part of the total $pn \to d\pi^+\pi^-$ cross section in the
region of the dibaryon resonance $d^*(2380)$. Blue filled  circles represent the
experimental results from WASA-at-COSY \cite{d+-,d00}, the horizontal bars give
the binning width used. Red solid and dotted curves show the calculations of
Ref.~\cite{EO}. From Ref. \cite{Comment}.
}
\label{fig2}
\end{figure}

In Ref. \cite{EO} the results of the two-step calculations were not confronted with experimental data. Hence we 
display both the calculations and the WASA-at-COSY data \cite{d+-} in Fig.~\ref{fig2} for the isoscalar part of the $d\pi^+\pi^-$
channel. Despite of tuning the fit on the cross section for the step-1 process the calculated peak structure comes out too low in energy by 30 - 40 MeV,
which is far outside experimental uncertainties \cite{d+-,d00}.
The fact that the peak calculated for the $d\pi^+\pi^-$
channel misses the measured peak by about 40 MeV is
associated in Ref. \cite{EO} with a pretended experimental resolution of 20 MeV in $\sqrt{s}$. However, here the authors of Ref.
\cite{EO} mix up the experimental resolution with the bin width
used for the presentation of differential distributions in
Ref. \cite{d+-}. Furthermore, a finite experimental energy resolution affects the width of a resonance structure, but
not its position. The binning used for the presentation of total cross section was 10 MeV in Refs. \cite{d00,d+-} — see
Fig. 2 — and the high precision COSY beam had a resolution in the sub-MeV range.

Since no specific angular momenta are considered in the formalism of
Ref. \cite{EO}, the calculated structure in the final channel contains a
priori a variety of spin-parity combinations. From partial-wave analyses
\cite{SAIDpid,SAIDpppid} of the step-2 reaction, the $pp \to d\pi^+$ reaction, 
we know that 62$\%$ of its 
total cross section is due to the $^1D_2$ partial wave between the incident
proton pair (which leads to a $^3S_1 - ^3D_1$ nucleon pair in the outgoing
channel associated with an emerging pion in relative $P$-wave, often
abbreviated as $^1D_2P$ partial-wave channel). Therefore the
proton pair emerging from the step-1 reaction, {\it 
 i.e.} the $pp(I=0)\to pp\pi^-$ reaction, should be predominately just in this
$^1D_2$ partial wave, in order to transport most part of the total step-1 reaction
cross section to the step-2 part and form a structure with $I(J^P) =
0(3^+)$. However, the $^1D_2$ partial wave between 
the emerging protons is only marginal if existent at all, as has been
demonstrated in a recent partial-wave analysis \cite{ASnew} of both the $pp
\to pp\pi^0$ and the $np \to pp\pi^-$ reaction.

\begin{figure} 
\centering
\includegraphics[width=0.89\columnwidth]{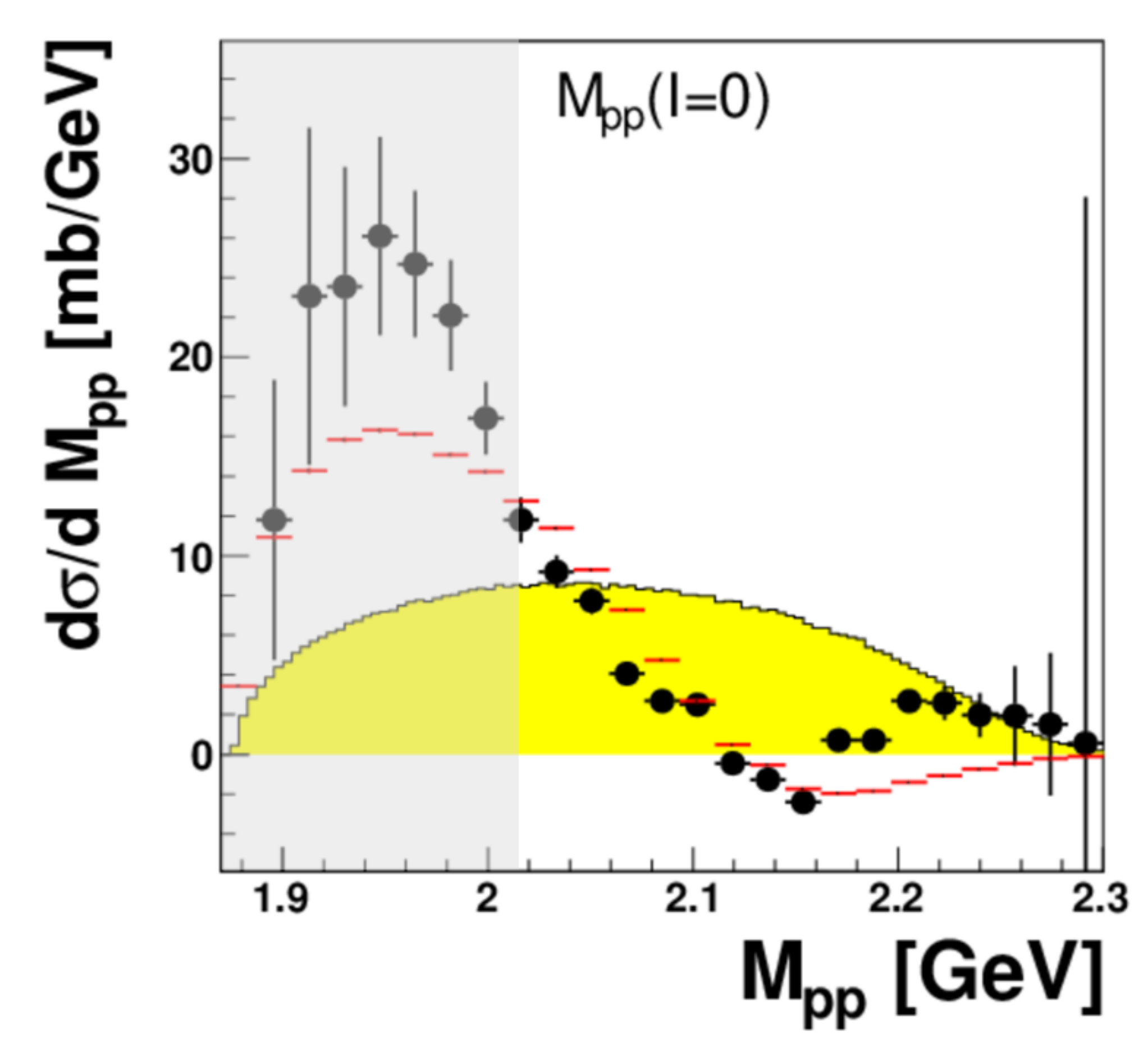}
\caption{\small 
  Isoscalar $pp$ invariant mass spectrum $M_{pp}(I=0)$ obtained from the
  difference of the corresponding distributions in the $pp \to pp\pi^0$ and
  $pn \to pp\pi^-$ reactions by use of eq.~(1) in
  Refs. \cite{NNpi,NNpicorr,NNpire}. The phase-space distribution is
  indicated by the (yellow) shaded region. The red dashed histogram gives a
  conventional $t$-channel calculation for Roper excitation
  \cite{NNpi,Luis}. The gray shaded region is below the threshold of 2.015 GeV
  for the $pp \to d\pi^+$ reaction, {\it i.e.} not available for the step-2
  reaction. 
}
\label{fig3}
\end{figure}

\begin{figure} 
\centering
\includegraphics[width=0.99\columnwidth]{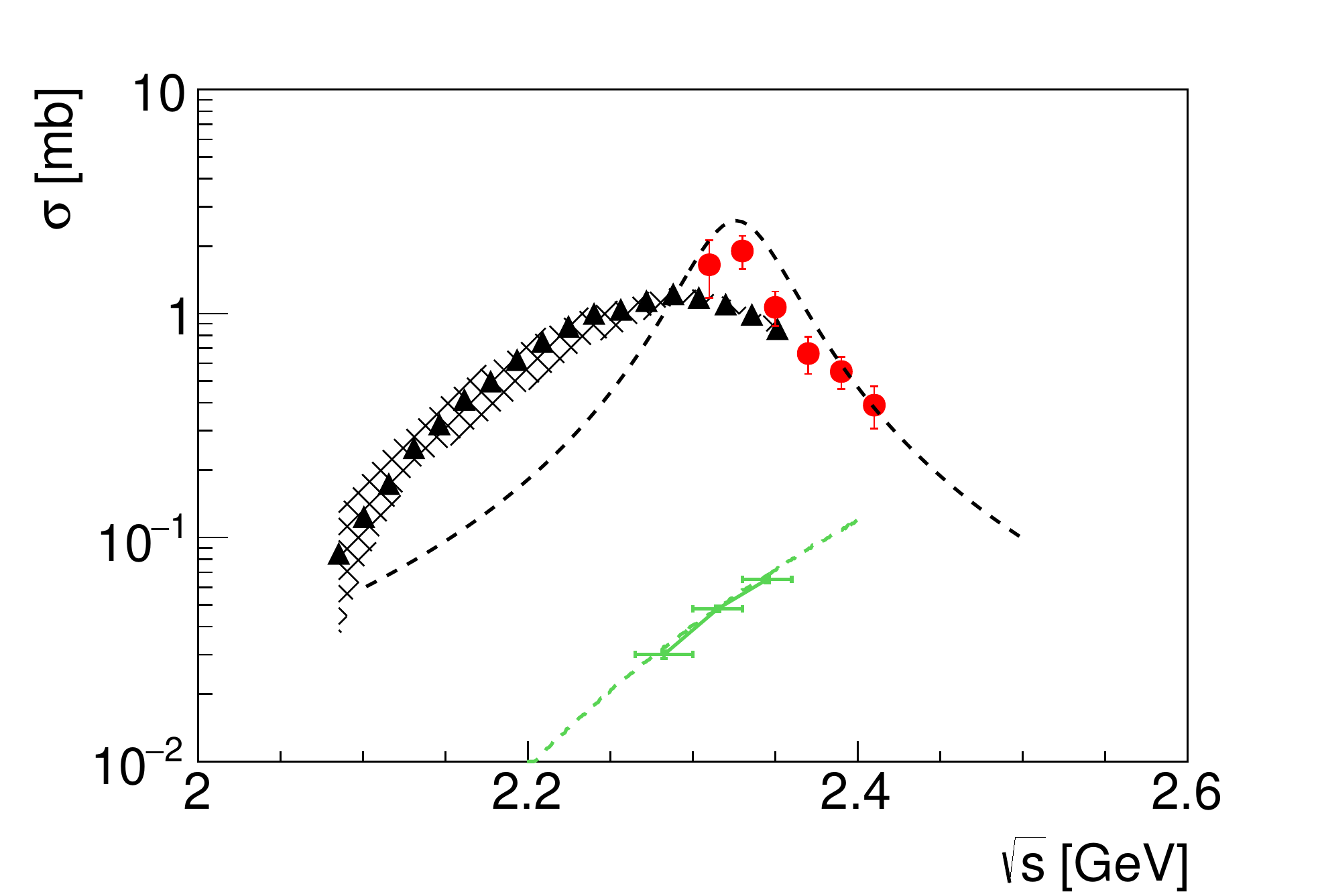}
\includegraphics[width=0.99\columnwidth]{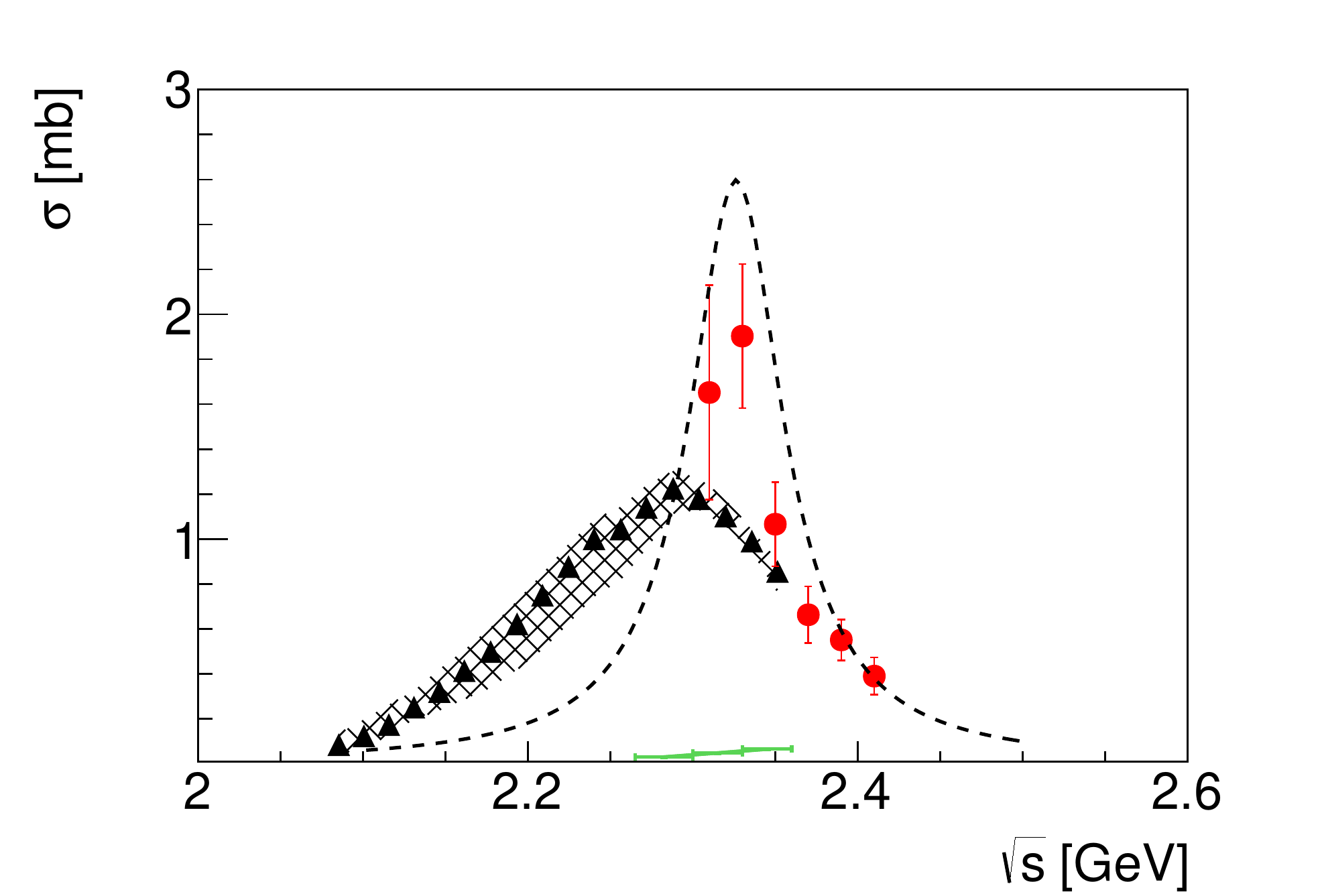}
\caption{\small  
Energy dependence of the isoscalar cross section for single-pion production in $NN$ collisions in logarithmic (top) and linear (bottom) scale. Shown are the experimental results from
WASA-at-COSY \cite{NNpi,NNpicorr,NNpire} (solid red dots) as well as the
results of partial-wave analyses of Ref. \cite{AS} (hatched black band)
The dotted curve shows the fit of Ref. \cite{EO}, the green markers
represent the partial-wave analysis results of Ref. \cite{ASnew} for the
$np(^3D_3) \to pp(^1D_2)\pi$ contribution with a fit curve shown as a green dashed line.
}
\label{fig4}
\end{figure}

The finding of this partial-wave analysis \cite{ASnew} is in accord with the isoscalar
proton-proton invariant-mass 
spectrum deduced from the WASA experiment \cite{NNpi,NNpire}, which is
displayed in Fig.~\ref{fig3}. As we can see there, the strength is concentrated just at
lowest $pp$-masses. Approximately $2/3$ of the strength is situated below
2.105 GeV, which is the threshold for the step-2 reaction, the $pp \to d\pi^+$
process. Hence only $1/3$ of the total step-1 reaction cross section is
kinematically available for the step-2 reaction. In addition we know from the
partial-wave analyses results for isoscalar single-pion production \cite{ASnew,AS} that
there are practically only $S$- and $P$-waves between the proton pair emitted
from the step-1 reaction. The $pn(^3D_3) \to pp(^1D_2)\pi$ partial wave contributes only with a few percent to the total cross section of the step-1 reaction, as may be seen in Fig.~\ref{fig4}, where the results of the partial-wave analysis of Ref.~\cite{ASnew} for this partial wave are indicated by the green horizontal bars.
Taking these facts into account, the
calculated two-step cross section for the $d\pi^+\pi^-$ channel drops by
nearly two orders of magnitude -- already on a qualitative level.

As has been demonstrated in a recent publication \cite{NNpire}
the world pool of data
\cite{NNpi,NNpicorr,NNpire,AS,Sarantsev,Rappenecker,Tsuboyama,Dakhno} for
isoscalar single-pion production does not 
support the Breit-Wigner fits of Ref. \cite{EO} of having a peak at at 2.33 -
2.34 GeV with a width of 70 - 80 MeV, but rather supports the
result of Ref. \cite{NNpi} of having a peak at 2.31(1) GeV and a width of
150(10) MeV. Fig.~\ref{fig4} shows the energy dependence of the isoscalar single-pion
production cross section based on data from WASA-at-COSY (solid dots)
\cite{NNpi,NNpicorr,NNpire} and partial-wave analysis results from
Ref.\cite{AS} (hatched band) together with the fit of Ref. \cite{EO} (dotted
curve) and the partial-wave analysis results of Ref. \cite{ASnew} for the
$np(^3D_3 \to pp(^1D_2)$ contribution (horizontal bars).
We note in passing that the excursion of the WASA-at-COSY data point at $\sqrt
s$= 2.32 GeV seen in Fig.~\ref{fig4} 
and which was focused on in Ref. \cite{EO} could, indeed, suggest a tiny
narrow structure on top of the broad isoscalar Lorentzian. However, the
neighboring data points are low -- both those from WASA and from other
experiments \cite{Sarantsev,Rappenecker,Tsuboyama}. Hence the 3 $\sigma$
excursion at $\sqrt s$ = 2.32 GeV appears to be of no particular significance
as discussed in more detail in Ref. \cite{NNpire}.


\begin{figure} 
\centering
\includegraphics[width=0.99\columnwidth]{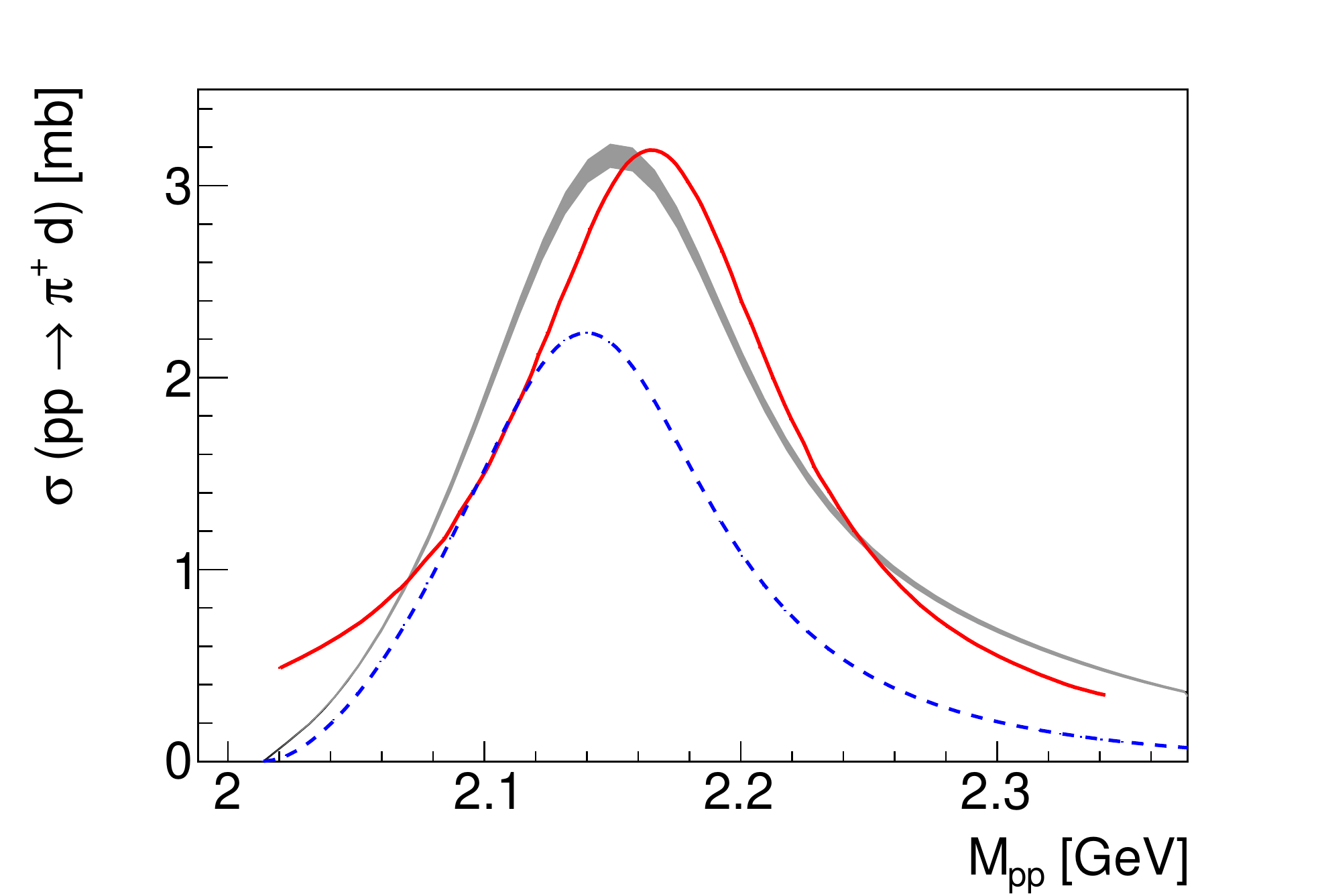}
\caption{\small  
Energy dependence the total $pp\to d\pi^+$ cross section plotted in dependence of the $pp$-invarinat mass $M_{pp}$. Shown are the solution of the SAID partial-wave analysis (grey) \cite{SAID}, the parametrisation of Ref.~\cite{EO} (red) as well as the SAID solution for the single  $^1D_2P$ partial wave (blue dashed).
}
\label{ppdpi}
\end{figure}

In conclusion we find that the two-step ansatz of Ref. \cite{EO} is far away
from giving any explanation for the $d^*(2380)$ peak in double-pionic
fusion. But we may ask what kind of prediction delivers the two-step ansatz,
if we feed it with correct experimental information. To parametrise the $pp\to
d\pi^+$ reaction the authors of Ref.~\cite{EO} took very old data from
Ref.~\cite{RS} with large error bars. This reaction has been studied meanwhile
in details~\cite{SAIDpid}, so the database contains more than thirty thousand
points leading to extremely small uncertainties in the total
cross-section. The SAID partial wave analysis \cite{SAID} claims a 2\% error
\cite{Igor} for an energy dependent solution. As we can see from Fig.~\ref{ppdpi} the fit of Ref.~\cite{EO} has substantial deviations with regard to the latest cross-section parametrisation. In order to perform a calculation of the sequential process cross-section, one should, however, not use the total cross-section, but only the cross-section of the $^1D_2P$ partial wave. Though the $^1D_2P$ partial wave is dominant in the $pp\to d\pi^+$ reaction, it is far from covering 100\% of the total cross-section and also substantially different from the Ref.~\cite{EO} parametrisation. Ideally, to calculate the sequential process correctly, one would need to redo an integration with a proper parametrisation. For simplicity we will just account for the difference in strength by 
\begin{equation}
    \frac{\int_{2015}^{2300} \sigma(pp\to d\pi^+)^{^1D_2}}{\int_{2015}^{2300} \sigma(pp\to d\pi^+)^{total}}=0.62,
\end{equation}
where the index $^1D_2$ refers to the $^1D_2$ partial wave in the initial $pp$ channel.

\begin{figure} 
\centering
\includegraphics[width=0.99\columnwidth]{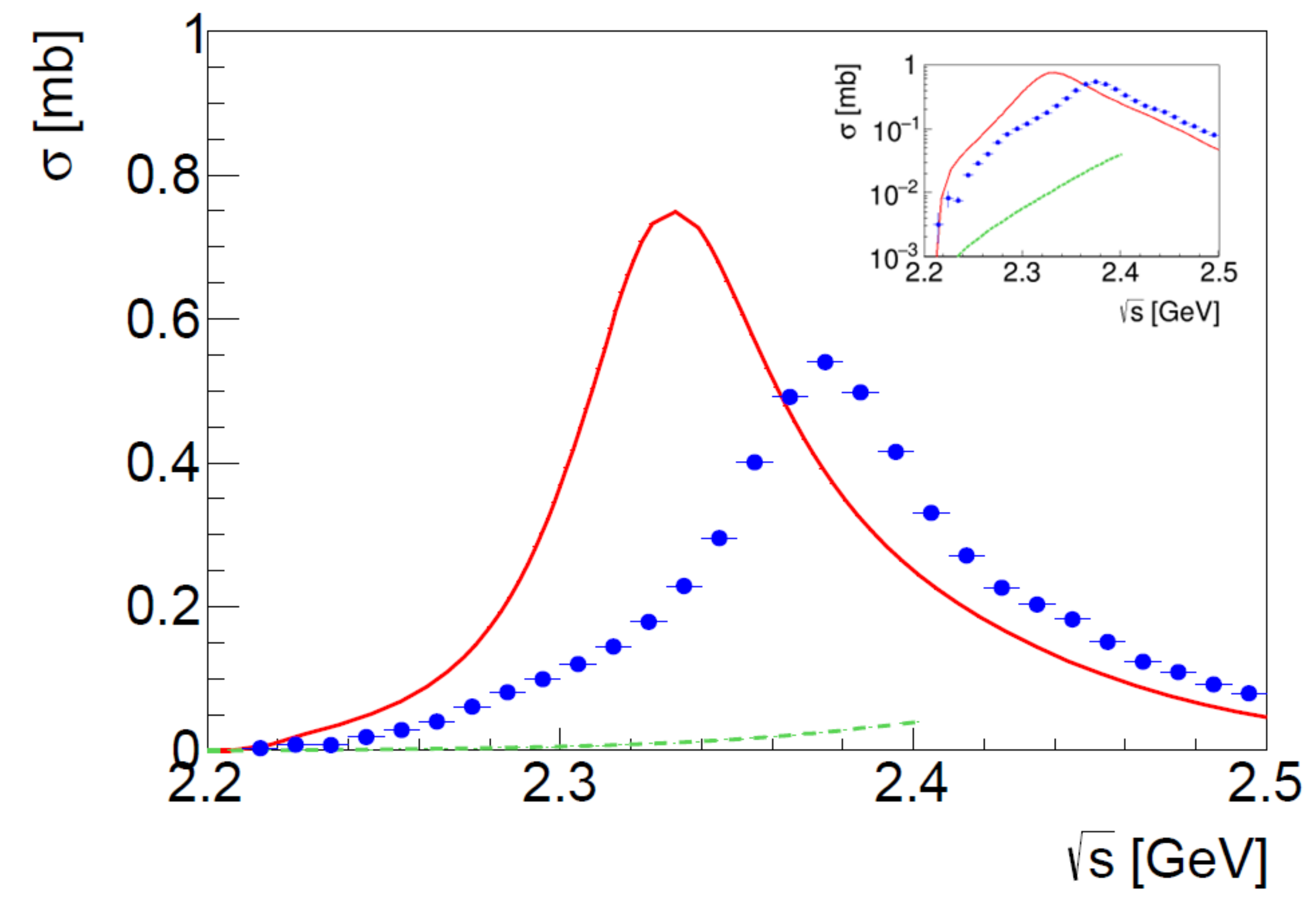}
\caption{\small  
The same as Fig.~\ref{fig2}, but with the additional green dashed curve, which
shows the corrected calculation, where the $pn(I=0)\to NN\pi$ cross-section of
Ref.~\cite{EO} was substituted by the PWA extracted $pn(^3D_3)\to
NN(^1D_2)\pi$ cross-section of Ref~\cite{ASnew} and the $NN(^1D_2)\to d\pi^+$
cross section was properly scaled, see Fig.~\ref{ppdpi}. The inset shows the same distribution in log scale.
}
\label{fig6}
\end{figure}

From the partial-wave
analysis of Ref. \cite{ASnew} we know the proper cross section $\sigma (pn(^3D_3)\to NN(^1D_2)\pi)$. So we can unfold the $\sigma(pn\to d\pi\pi)$ prediction of Ref.~\cite{EO} from its unreasonably narrow Lorentzian for the isoscalar single-pion total cross-section $\sigma(NN\pi)_{I=0}^{narrow}$ and replace it by the correct cross-section of the proper partial wave, which can mimic the $d^*(2380)$ resonance. In a simplified manner this can be accomplished by
\begin{equation}
\begin{split}
\sigma(d\pi\pi)^{cor} = & \sigma(d\pi\pi)^{original}\cdot \\  & \cdot\frac{\sigma(pn(^3D_3)\to NN(^1D_2)\pi)}{\sigma(NN\pi)_{I=0}^{narrow}}\cdot 0.62.
\end{split}
\end{equation}
The result of such unfolding is presented in Fig~\ref{fig6} by the green dashed line. For the cross section $\sigma(pn(^3D_3)\to NN(^1D_2)\pi)$ we took a fit of the Ref.~\cite{ASnew} data based on a phase-space distribution weighted with the fourth power of the beam momentum,  $(P_{beam}^{CMS})^4$, in order to account for the $D$-wave dependence. In Fig.~\ref{fig4}  the fit is shown by the green dashed line. For the range of interest $\sqrt{s}\in[2.25,2.4]$~GeV the fit seems to be reasonable.

As can be seen in Fig.~\ref{fig6} the removal of the unreasonably narrow Lorenzian for the isoscalar single-pion production channel automatically removes the peaking from the $\sigma(d\pi\pi)$ cross-section. A smoothly rising $\sigma(pn(^3D_3)\to NN(^1D_2)\pi)$ dependence replicates itself in the $\sigma(d\pi\pi)$ cross-section. The overall contribution of such a process is in the order of 5\% for the $d^*(2380)$ peak or smaller. However, one needs to be cautious here and consider this prediction as an upper limit, since it still does not account for the exact $M_{pp}$ distribution for a $(pn(^3D_3)\to NN(^1D_2)\pi)$ partial wave. Also its cross-section cannot rise infinitely for higher energies and should be damped. As can be seen on Fig.~\ref{fig4}, such effects should be sizeable already at 2.4~GeV. 

Whereas the two-step process cannot reproduce the $d^*(2380)$, a resonance in a $^3D_3$ partial wave, we may ask if it can give any meaningful contributions in the other partial waves. From the partial-wave analysis of Ref. \cite{AS} we know that the $^3S_1-^3D_1$ and $^1P_1$
partial waves in the initial $pn$ system dominate the $pn(I=0) \to pp\pi^-$
reaction with the consequence of having only $S$- and $P$-waves between the
emerging proton pair. From the partial-wave analyses of
Refs. \cite{SAIDpid,SAIDpppid} we learn that initial $S$- and $P$-waves
contribute to only about 10$\%$ to the total $pp \to d\pi^+$ cross
section. Putting these facts together with the Breit-Wigner fit result of
$\sigma_{peak}$ = 1.4 mb (instead of 2.6 mb), m = 2.31(1) GeV (instead of 2.33
- 2.34 GeV) and width $\Gamma$ = 150(10) MeV (instead of 70 - 80 MeV), we
obtain with the two-step ansatz again a prediction for a 
resonance-like structure, but now around 2.32 GeV with a width of about 150
MeV and a peak cross section of about 0.06 mb. If we in addition account for
the fact that only $1/3$ of the strength in the $M_{pp}$ spectrum is above
the threshold for the step-2 reaction, then the peak cross section from the
two-step ansatz decreases further to about 0.02 mb.

\begin{figure} 
\centering
\includegraphics[width=0.99\columnwidth]{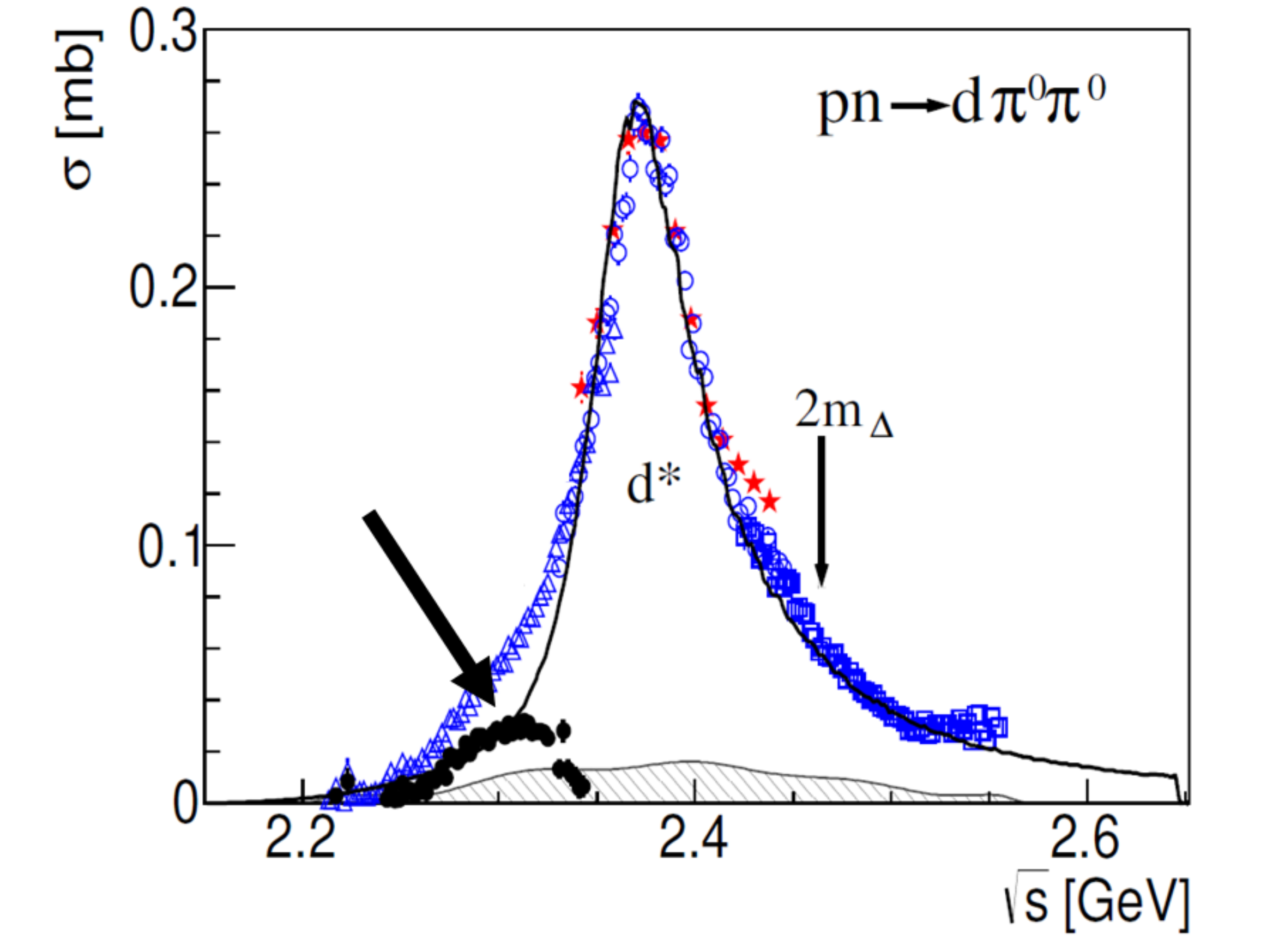}
\caption{\small 
Energy dependence of the total cross section for
  the $pn \to d\pi^0\pi^0$ reaction as measured by WASA-at-COSY. The blue open
  symbols represent the data of Ref.~\cite{d00} normalized to the data
  (red stars) of Ref.~\cite{d+-}. The hatched area gives an estimate
  of systematic uncertainties. The solid curve displays a calculation of the
  $d^*$ resonance with momentum-dependent widths \cite{abc}. It includes both
  Roper and $\Delta\Delta$ $t$-channel excitations as background
  reactions. The black filled dots show the difference between data and this
  calculation in the low-energy tail of $d^*(2380)$. The thick arrow points to 
  the resulting bump structure. From Ref. \cite{NNpire}.
}
\label{fig7}
\end{figure}

Indeed, we may associate this prediction with a possible peak reported
\cite{NNpire} recently for the $d\pi^0\pi^0$ channel, which by isospin
relation has a factor two smaller cross section than the isoscalar part of the
$d\pi^+\pi^-$ channel. In Fig.~\ref{fig7}, which has been taken from
Ref. \cite{NNpire}, the experimental total cross section is displayed for the
$np \to d\pi^0\pi^0$ reaction. If a Breit-Wigner ansatz with a
momentum-dependent width \cite{abc} is used for the description of the
$d^*(2380)$ resonance, then the data on the low-energy side of the $d^*(2380)$
resonance are underpredicted. The difference between data and resonance
description yields a small bump around 2.32 GeV with a width of about 150 MeV
and a peak cross section of about 0.03 mb. i.e. an order of magnitude smaller
than the neighboring peak cross section of $d^*(2380)$. Amazingly, these
features fit very well to the prediction of the two-step ansatz, if fed
with the proper cross section data.

In Ref. \cite{NNpire} the small bump was interpreted as consequence of
possible dibaryon resonances with $I(J^P) = 0(1^+)$ and $0(1^-)$, which
produce the bump in isoscalar single-pion production due to the $^3S_1-^3D_1$
and $^1P_1$ partial waves between the incident $pn$ pair. From the fact that
the isoscalar proton-pion invariant-mass distribution exhibits strength only
in the region of the Roper excitation (see Fig. 6 in Ref. \cite{NNpi}) it was
concluded that these dibaryon resonances must have a $N^*N$ structure. And
since the Roper resonance decays by single- and by two-pion emission, this
structure must be present both in single- and double-pion production. In
Ref. \cite{NNpire} it was shown that the known branchings of the Roper decay
fit to the relative size of the bumps observed in isoscalar single- and
double-pion production. 

Actually, it is not surprising that the two-step ansatz of Ref. \cite{EO} fits
to the interpretation of Ref. \cite{NNpire}. By describing the isoscalar
single-pion production with a Breit-Wigner form the authors of Ref. \cite{EO}
implicitly assume a resonance in the isoscalar $pn$ system -- without stating
that explicitly. Thus the two-step ansatz simulates in essence at least part of the
dibaryon interpretation of Ref. \cite{NNpire} for the small bump in the
$d\pi^0\pi^0$ channel, if one keeps in mind that the upper leg of the two-step
graph after the first interaction blob can be reinterpreted as the sequential
Roper decay $N^* \to \pi \Delta \to \pi\pi N$.

Finally we would like to comment on the situation of $d^*(2380)$ in photon absorption on the deuteron, where according to Ref.~\cite{EO} a sequential process should not occur and hence would be decisive for the interpretation of the prominent peak in the $pn \to d\pi^0\pi^0$ reaction. According to the theoretical prediction of Ref.~\cite{IHEP} the branching $d^*(2380) \to d\gamma$ is only in the order of $10^{-5}$ and hence a detection of a signal in $\gamma d$-induced reactions is very difficult. It is true that the current experimental hints \cite{ELPH,MAMI,BGOOD} for $d^*(2380)$ in the $\gamma d \to d\pi^0\pi^0$ reaction are not yet conclusive due to smallness of total cross section and large backgrounds, though the differential $M_{\pi^0\pi^0}$ spectrum in Ref.~\cite{BGOOD} is in favor of a $d^*(2380)$ contribution. More conclusive results are expected to come from MAMI, where dedicated measurements of $\gamma d \to d\pi^0\pi^0$ will be performed with active deuteron target and deuterium-TPC setups. In the situation, where background due to conventional reaction processes is large, measurements of polarization observables are known to be very helpful. In $pn$ elastic scattering, where the contribution of $d^*(2380)$ to the total cross section is only marginal, it was demonstrated that polarization measurements reveal a pronounced effect of $d^*(2380)$ in the analyzing power, whereas its effect in the unpolarized differential cross section is small and quite unspecific~\cite{np,npfull,npel}. Similarly, evidence of $d^*(2380)$ photoexcitation is found in polarization observables of deuteron photodisintegration $\gamma d\to pn$~\cite{MBSigma,MBPy,MBCx}, however, more polarization measurements (e.g. T, E, F, G observables) and a partial wave analysis would be highly desirable to settle the question of the $d^*(2380)$ photoproduction.

\section{Summary}

In conclusion the two-step ansatz of Ref.~\cite{EO} has been shown to be not
able to provide an alternative interpretation of $d^*(2380$, if fed with
proper experimental data for the step-1 and step-2 cross sections and if the
constraints for proper partial waves are taken into account. However, we have
demonstrated that the two-step ansatz may give reasonable results for a
possible small, but broad structure below the $d^*(2380)$ peak, which was
recently interpreted as a consequence of a possible dibaryonic excitation in
isoscalar single-pion production.


\section{Acknowledgments}

We acknowledge
valuable discussions with E. Oset, A. Gal and I. Strakovsky.
This work has been supported by DFG (CL 214/3-3) and by the U.K. STFC ST/L00478X/2, ST/V001035/1.

\end{document}